\documentclass{PoS}

\usepackage{amsmath}
\usepackage{calligra}
\usepackage{calrsfs}
\DeclareMathAlphabet{\mathcalligra}{T1}{calligra}{m}{n}
\DeclareMathAlphabet{\pazocal}{OMS}{zplm}{m}{n}
\usepackage{notoccite}

\def\dm{\text{\tiny{DM}}}

\def\rh{\text{\tiny{RH}}}
\def\max{\text{\tiny{MAX}}}

\def\MP{M_{\rm \tiny{P}}}

\def\Ga{\Gamma}

\def\M{\pazocal{M}}

\def\Lag{\pazocal{L}}

\def\rad{\gamma}
\def\x{\text{\textit{\tiny{X}}}}

\def\a{\alpha}

\def\g{\gamma}
\def\s{\sigma}

\def\m{\mu}
\def\n{\nu}
\def\e{\epsilon}

\def\zp{\text{\tiny{$Z'$}}}
\def\xa{\text{\tiny{$X_1$}}}
\def\xna{\text{\tiny{$X_N$}}}

\def\gsm{g_{\text{\tiny{SM}}}}
\def\gdm{g_{\text{\tiny{DM}}}}
\def\mh{m_{\tilde h}}

\title{Freeze-in production of dark matter through spin-1 and spin-2 portals}

\ShortTitle{Freeze-in production of dark matter through spin-1 and spin-2 portals}

\author{\speaker{Ma\'ira Dutra}\\
        Carleton University\\
        E-mail: \email{mdutra@physics.carleton.ca}}

\abstract{
In this conference, I have talked about two scenarios in which the out-of-equilibrium production of dark matter (DM) particles in the early universe is unavoidable. In the first one \cite{bhattacharyya_freezing-dark_2018}, we extend the standard model (SM) of particle physics by an extra $U(1)$ gauge group under which all the SM particles are neutral. We then consider DM candidates interacting only with the new spin-1 gauge boson, a heavy $Z^\prime$. We assume the presence of heavy beyond the SM fermions charged under both extra $U(1)$ and SM $SU(3)_c$, allowing for a feeble connection between DM and gluons. In the second scenario \cite{bernal_spin-2_2018}, we assume that the interaction between DM and SM particles are only mediated by gravitons and massive spin-2 fields, being therefore suppressed by the Planck and some intermediate scales, respectively. In both models, we show that the SM particles are able to produce the right amount of DM candidates via freeze-in at most in the early stages of the radiation era, for DM mass in the range $10^{-3}-10^{14}$ GeV. We have shown that if heavy mediators were produced on-shell within a period of entropy production in the early universe, as in the post-inflationary reheating, the DM relic density may be enhanced by many orders of magnitude relative to the usual instantaneous reheating approximation.
}

\FullConference{XXIX International Symposium on Lepton Photon Interactions at High Energies - LeptonPhoton2019\\
		August 5-10, 2019\\
		Toronto, Canada}

\begin{document}

\section{Introduction}

The nature of dark matter (DM) particles is one of the biggest open problems in the interface of particle physics and cosmology. In the hope to detect DM, we assume that the dark and visible sectors are able to interact. Weakly (feebly) interacting massive particles, the WIMPs (FIMPs), are DM candidates whose interactions with the visible sector are strong (feeble) enough so that they were initially (never) thermalized. 
FIMPs were then produced from out-of-equilibrium decays or annihilations of species of the thermal bath in the early universe (EU), the \textit{freeze-in} mechanism \cite{chung_production_1998,hall_freeze-production_2010,bernal_dawn_2017}. Nowadays, direct detection (DD) experiments are already excluding many WIMP candidates \cite{arcadi_waning_2018} and, in the next decade, they might start probing FIMP candidates as they are becoming more and more sensitive \cite{hambye_direct_2018}. An important point, though, is that the couplings probed by DD experiments, which might be also responsible for an initial kinetic equilibrium situation, need to be smaller and smaller as to evade the bounds. This is, therefore, a clear potential problem for models invoking the freeze-out mechanism, easily leading to overproduced DM candidates. In this context, the freeze-in mechanism is a good motivation for the pursue of DD searches. 

Since FIMPs have an out-of-equilibrium origin, their final relic density might depend on initial conditions, i.e. on very high-energy physics taking place in the EU. If the freeze-in happens through contact interactions, for instance, it usually happens close to the FIMP mass, and the direct couplings need to be extremely small. In this case, the final relic density depends only on masses and couplings. In the presence of heavy mediators or effective couplings, a good relic density can be achieved for larger dimensionless couplings, the freeze-in depends on initial conditions and happens earlier, at the scale of the heavy mediator mass. In this context, given our ignorance about the scale at which an eventual reheating epoch would have happened in the EU, it is important to take into account the possibility that the mass scales of particles involved in the freeze-in process are close to a reheating scale. The appealing aspect of such a possibility is that heavy fields appear in many well-motivated extensions of the standard model (SM). In what follows, we consider two natural realizations of the freeze-in mechanism, in the sense that very feeble couplings between the dark and visible sectors are inevitable. 

\section{Freeze-in during reheating}\label{sec:FIrh}

The Boltzmann fluid equation tells us that the total number of DM particles ($N_\dm = n_\dm a^3$) can only change in the expanding volume of the universe ($a^3$) if they are globally produced or annihilated, and this information is encoded in the reaction rate density $R_\dm$. It reads $\frac{dN_\dm}{dt}=R_\dm a^3$. The way the scale factor changes with time depends on the total energy density of the universe ($\rho(t)$), $d\ln a/dt = H(t) \propto \sqrt{\rho (t)}$, with $H(t)$ the Hubble expansion rate. If $\rho(t)$ is a known function of the temperature of the SM thermal bath ($T$), provided that we know which kind of species dominates the expansion rate, we can rewrite the Boltzmann fluid equation in terms of the yield of DM $Y_\dm \equiv N_\dm/S$, with $S$ the total entropy in a comoving volume, and study how $Y_\dm$ changes with time ($1/T$). However, if $\rho(t)$ varies during a given period, provided that some entropy is being injected, the yield of DM after such a period gets diluted by the ratio of entropy after and before. This dilution might be the result of the decay of an unstable field ($\phi$) into particles of the thermal bath ($\gamma$). This process of energy injection is what we call "reheating". In the context of inflationary frameworks, the inflaton field would make the universe to expand exponentially and then decay into SM species, producing a huge amount of entropy. This is referred to as the inflationary reheating. Therefore, during reheating we need to solve the DM evolution along with the evolution of the unstable field and its decay products. 

\begin{figure}[t]
\centering
\includegraphics[scale=0.2]{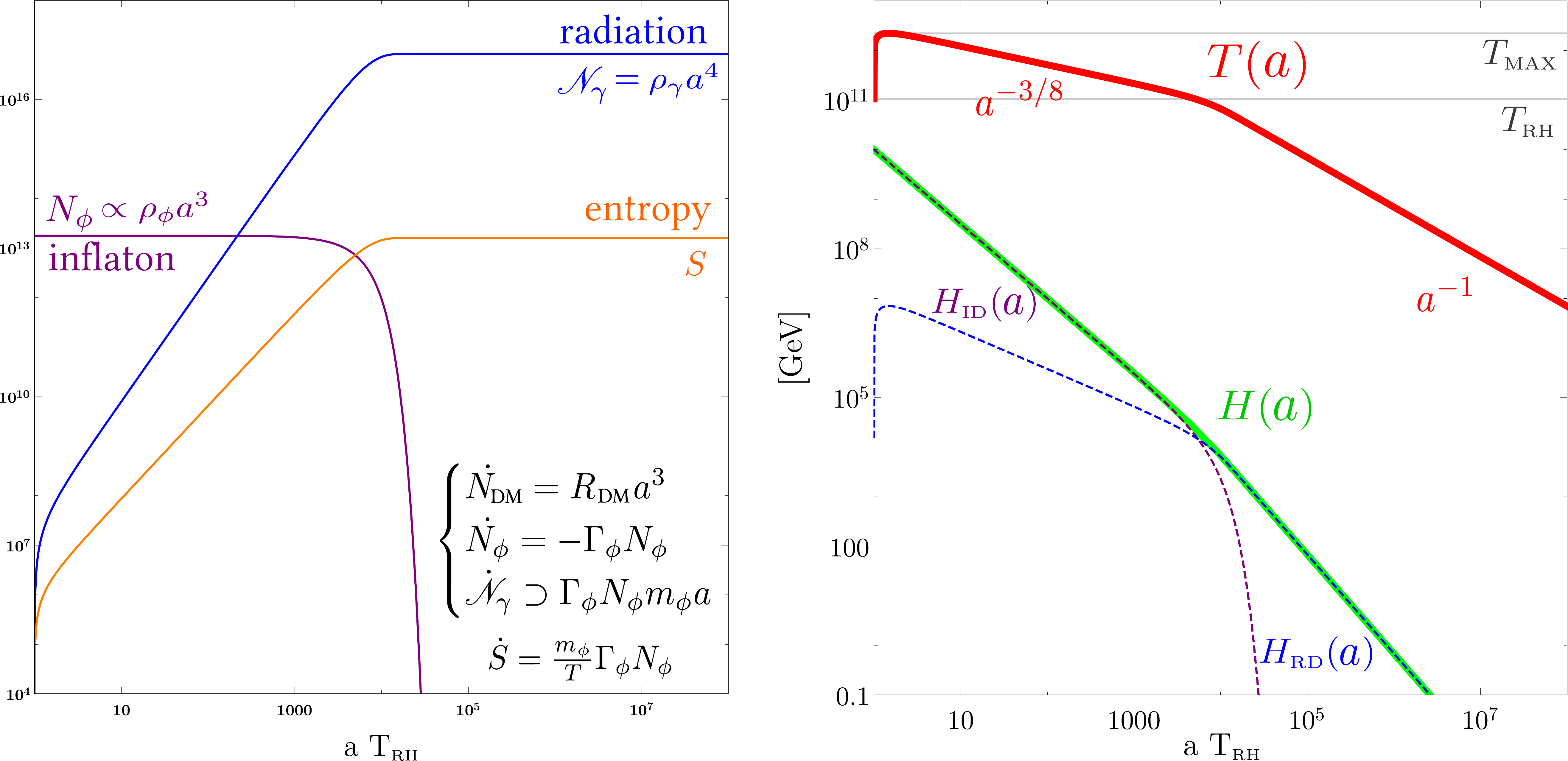}
\caption{\em \small Evolution of the inflaton-radiation system as function of the scale factor ($a$). }
\label{fig:inflationaryRH}
\end{figure}

In the left panel of Fig.\ref{fig:inflationaryRH}, we see the solutions for total numbers of inflaton and radiation as function of the scale factor normalized to the reheating temperature. We see that while inflaton is decaying, there is production of radiation, and therefore of entropy, until the total production of the radiation relic, around the so-defined reheat temperature ($T_\rh$). As we see, this is not an instantaneous process and DM could start being produced at temperatures above $T_\rh$. In the right panel of Fig.\ref{fig:inflationaryRH}, we see how $T$ evolves while this is happening (red curve). There is a maximal temperature for the thermal bath ($T_\max$) which might be much larger than $T_\rh$, depending on the initial conditions provided by the inflationary model \cite{giudice_largest_2001}. We also see how the $H(t)$ evolves (green curve) from an epoch of inflaton-domination (ID), between $T_\max$ and $T_\rh$, to an isentropic epoch of radiation-domination (RD). The relic density of DM have therefore two contributions \footnote{For a pedagogical derivation of the expression above, as well as of the set of Boltzmann fluid equations, we refer the reader to Ref. \cite{dutra_origins_2019}.}:
\begin{equation}\begin{split}\label{relicsplit}
\Omega_\dm^0 h^2 
&= \frac{m_\dm}{2.16 \times 10^{-28}} \left( \int_{T_0}^{T_\rh} dT \frac{g_s^*}{g_s \sqrt{g_e}} \frac{R_\dm (T)}{T^6} + 1.6\,c\,B_\rad  \,g_\rh^{-3/2}\, T_\rh^7 \int_{T_\rh}^{T_\max} dT g_e^* \frac{R_\dm (T)}{T^{13}} \right) \, .
\end{split}\end{equation}

Now, we might ask when does the freeze-in happens. Moreover, when do we need to carry about such a reheating period even if we believe that it happened before the DM genesis? 
In most of the cases, the squared amplitudes of our model will tell us what is the T-dependence of the reaction rate densities, and this is precisely what we need to know in order to find the answer. It is easy to see from Eq. \ref{relicsplit} that if $R_\dm \propto T^n$ for $n<5$, the RD production would happen at the lightest scale available and if $n>5$, the RD production happens at most at the reheating scale. The same analysis holds for the ID production, in which case that power of T is $n = 12$. Most DM models do not have strong enough T-dependence and the term accounting for the ID production is usually neglected.
However, in the low-energy limit of many structural extensions of the SM, we can have effective operators connecting the dark and visible sectors. In this context, since the T-dependence of the production rates are usually high, the production during a reheating process is actually crucial for a correct prediction of the relic density of FIMPs. In what follows we will see two examples of that.

\section{A heavy spin-1 portal}\label{spin1}

Extending the gauge group of the SM by an extra U(1) symmetry is a simple and phenomenologically rich option because an extra gauge boson, $Z^\prime$, could be produced at colliders. $Z^\prime$ bosons with masses of the order of a few TeV can easily be a portal for WIMPs in the hundred GeV to few TeV mass scale, especially if the WIMPs are Majorana fermions \cite{arcadi_waning_2018}. Stronger bounds on $Z^\prime$ by colliders, though, means they are heavier or interact more feebly, which closes the portal to WIMPs. 

In Ref. \cite{bhattacharyya_freezing-dark_2018}, we studied the case in which only extra heavy fermions ($\Psi_i$) are charged under the SM $SU(3)_c$ and $U(1)^\prime$ groups, inducing effective interactions of three vectors dubbed Generalized Chern-Simons terms \cite{anastasopoulos_anomalies_2006}. Our FIMP candidates, a fermion ($\chi$), an Abelian ($X_1$) and a non-Abelian ($X_N$) vectors, interact with $Z^\prime$ through a gauge-like and GCS terms respectively. Our effective Lagrangian reads
\begin{equation}\label{Eq:leffG}
\Lag_\text{eff} = \frac{1}{\Lambda^2} \partial^\a Z'_\a \e^{\m \n \rho
  \s}\text{Tr}[ G^a_{\m \n} G^a_{\rho \s}] +
\begin{cases} 
\alpha ~\bar{\chi}\g^\m \g_5 \chi Z'_{\mu} \\
\beta ~ \e_{\m\n\rho\s} {Z^\prime}^\m X_1^\n X_1^{\rho\s} \\
\g ~\partial^\a Z'_\a \e_{\m \n \rho \s}\text{Tr}[ X_N^{\m \n} X_N^{\rho \s}]
\end{cases} \, .
\end{equation}

We have four free parameters: the FIMP and $Z^\prime$ masses ($m_\chi, m_\xa ~\text{or}~ m_\xna$ and $M_\zp$), the scale of new physics $\Lambda$ and the overall coupling between the FIMPs and $Z^\prime$ ($\alpha, \beta ~\text{or}~ \gamma$). $\Lambda$ is the cut-off of the theory, above which an effective approach is not valid. It therefore needs to be above $T_\max$. $\alpha$ would be an order unity gauge coupling, $\beta$ might be small since it will be given in terms of the charges and gauge couplings of the heavy fermions, and $\gamma$ will be extremely small since it is essentially the squared inverse of a new physics scale, fixed for simplicity to $\Lambda$.

The limiting case of the squared amplitudes for gluon annihilations into DM feature non-resonant exchanges of $Z^\prime$:
\begin{equation}\begin{split}
\int d\Omega^*_{13} |\M|_\chi^2 = 2^{10} \pi
~\frac{\alpha^2}{\Lambda^4} \frac{m_\chi^2 }{M_\zp^4}
\frac{s^3(s-M_\zp^2)^2}{(s-M_\zp^2)^2+M_\zp^2 \Gamma_\zp^2}
\approx 2^{10}\pi ~\frac{\a^2}{\Lambda^4} \frac{m_\chi^2}{M_\zp^4}s^3 \,,
\label{Eq:m1}
\end{split}\end{equation}
\begin{equation}\begin{split}
\int d\Omega^*_{13} |\M|_{\xa}^2 = 2^{10} \pi ~
\frac{\beta^2}{\Lambda^4} \frac{s^3}{M_\zp^4} \frac{ (s-4m_{\xa}^2)
  (s-M_\zp^2)^2}{(s-M_\zp^2)^2+M_\zp^2 \Gamma_\zp^2} \approx
2^{10}\pi ~ \frac{\beta^2}{\Lambda^4} \frac{1}{M_\zp^4}s^4
\label{Eq:m2}
\end{split}
\end{equation}
and
\begin{equation}\begin{split}
\int d\Omega^*_{13} |\M|_{\xna}^2  = 2^{12} \pi~
\frac{\g^2}{\Lambda^4} \frac{s^5}{M_\zp^4} \frac{
(s-4m_{\xna}^2)
(s-M_\zp^2)^2}{(s-M_\zp^2)^2+M_\zp^2 \Gamma_\zp^2} \approx 2^{12}\pi ~ 
\frac{\g^2}{\Lambda^4}\frac{1}{M_\zp^4}s^6 \,,
\label{Eq:m3}
\end{split}
\end{equation}
where $s$ is the Mandelstam variable and $\Gamma_\zp$ is the decay width of $Z^\prime$.

For the purposes of qualitative understanding, the power $k$ of $s$ is related to the power of temperature in the rate ($n$) by $n=2k+4$. Notice the increasing T-dependence of the rates due to more and more effective operators connecting the dark and visible sectors: $R_\dm \propto T^{10}, T^{12}, 10^{16}$. 

\section{A heavy spin-2 portal}\label{spin2}

Another case where the freeze-in mechanism is expected to happen is inspired in theories seeking a unified description of the fundamental interactions. In a quantum description, gravity is sourced by stress-energy tensors $T^{\mu \nu}$ and thought to be mediated by massless spin-2 fields, $h_{\mu \nu}$, the graviton. In this way, all the SM particles would be very feebly coupled to gravitons. Since all the evidence for DM come from gravitational interactions, it is important to investigate whether the very feeble couplings of spin-2 fields to dark and visible matter would be enough for the freeze-in.
 
In our minimal model \cite{bernal_spin-2_2018}, we have a graviton, a massive spin-2 field ($\tilde h$) that appears in extra-dimension frameworks and a FIMP candidate, which can be a scalar, a fermion or a vector (denoted by $X$). Our effective Lagrangian reads
\begin{equation}\label{lagSpin2}
\Lag \supset \frac{1}{2\MP} h_{\m\n} \left(\sum_{i=SM} T^{\m\n}_i + T^{\m\n}_X \right) + \frac{1}{\Lambda} \tilde h_{\m\n} \left(\gsm \sum_{i=SM} T^{\m\n}_i + \gdm T^{\m\n}_X\right) \,.
\end{equation}

Any interaction with $h_{\m\n}$ would be universally Planck-suppressed, while the interactions with $\tilde h_{\m\n}$ might be much less suppressed, by some intermediate scale $\Lambda$. Our FIMP candidates would be produced from all the SM particles through s-channel exchange of massless and massive spin-2 fields.

The squared amplitudes in this case reads
\begin{equation}
\int d\Omega^*_{13} |\M_{ij}|^2 = \frac{\pi s^2}{60 \MP^4} f_{ij}^h (s,m_\x) + \frac{\pi \gsm^2 \gdm^2}{15 \Lambda^4} \frac{s^4}{(s-\mh^2)^2 + \mh^2 \Ga_{\tilde h}^2}  f_{ij}^{\tilde h} (s,m_\x) \,,
\end{equation}
where the functions $f_{ij}^{h, \tilde h}$, defined in Ref. \cite{dutra_origins_2019}, are just numeric factors in the limit $s \gg m_\dm^2$.

Notice that graviton exchange leads to $R_\dm \propto T^8$, while the T-dependence of $\tilde h$ exchange depend on the ratio $m_{\tilde h}/\sqrt{s} \sim m_{\tilde h}/T$, being $T^8, T K_1(m_{\tilde h}/T)$ or $T^{12}$ corresponding to a light, resonant or heavy regime.

\section{Agreement with relic density constraints: comparative results}\label{results}

\begin{figure}[t]
\centering
\includegraphics[scale=0.3]{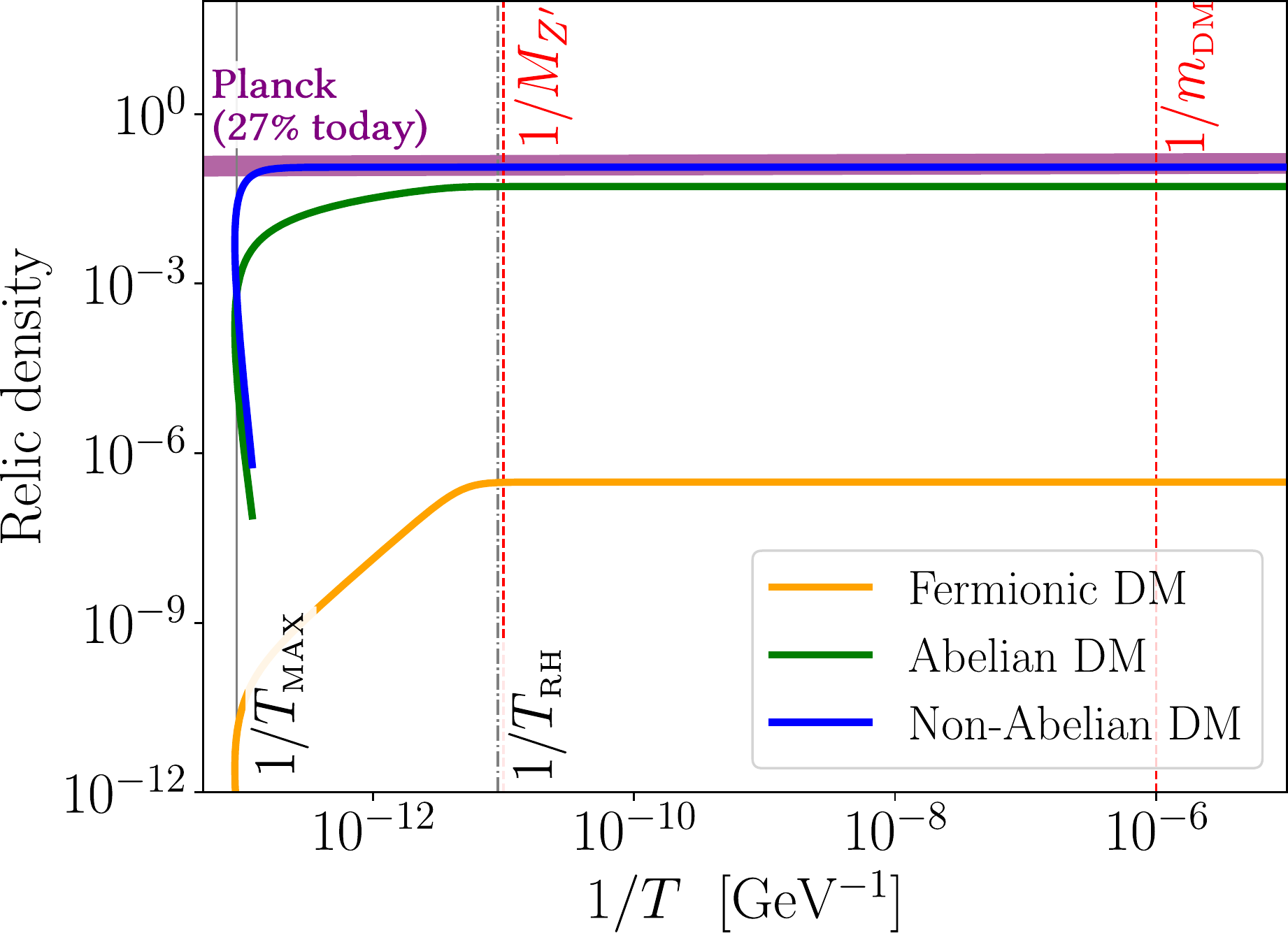}
\hspace{.3cm}
\includegraphics[scale=0.3]{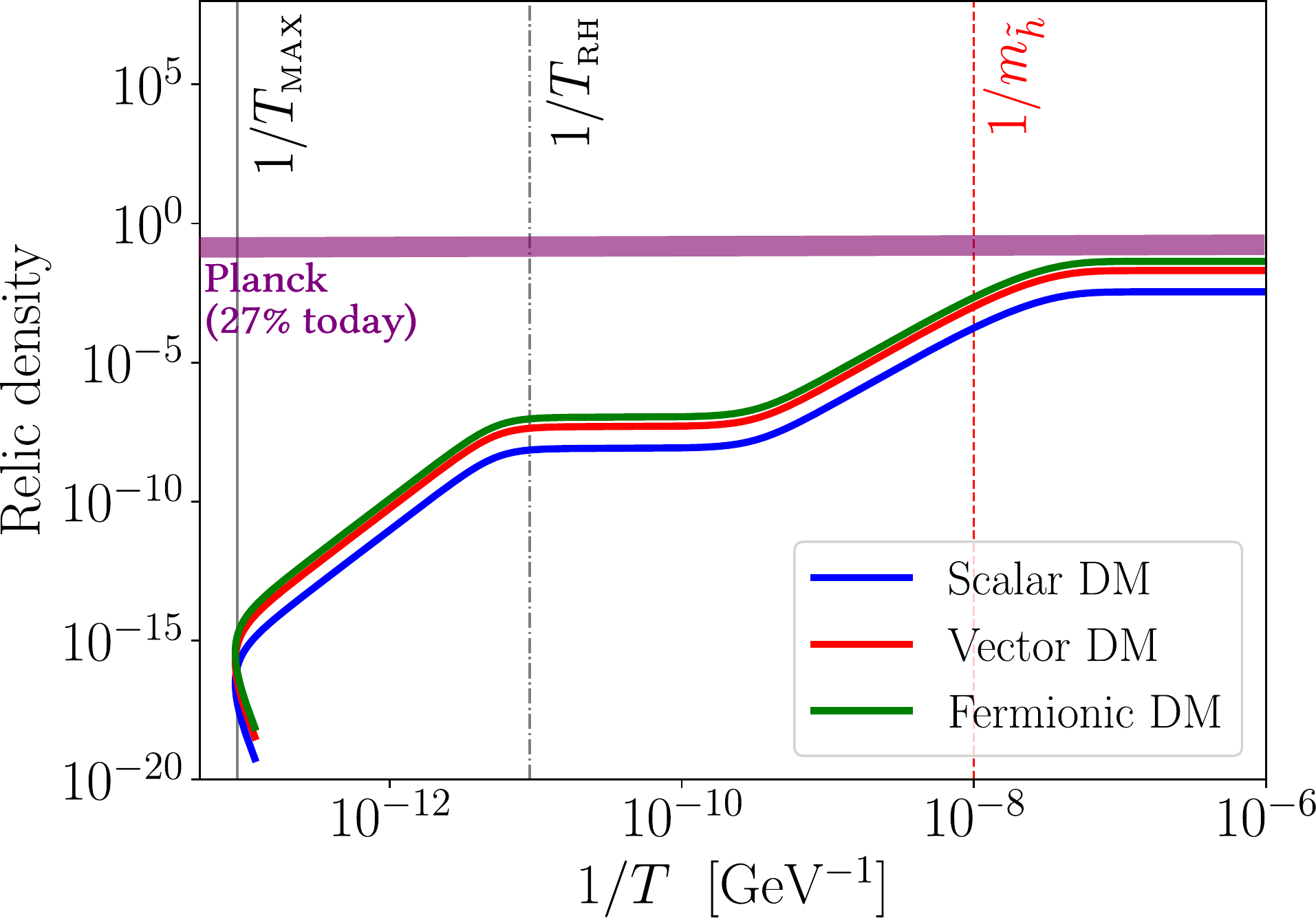}
\caption{\em \small Evolution of the relic density of our FIMP candidates in the spin-1 (left) and spin-2 (right) portals.}
\label{fig:evolution}
\end{figure}

\begin{figure}[t]
\centering
\includegraphics[scale=0.3]{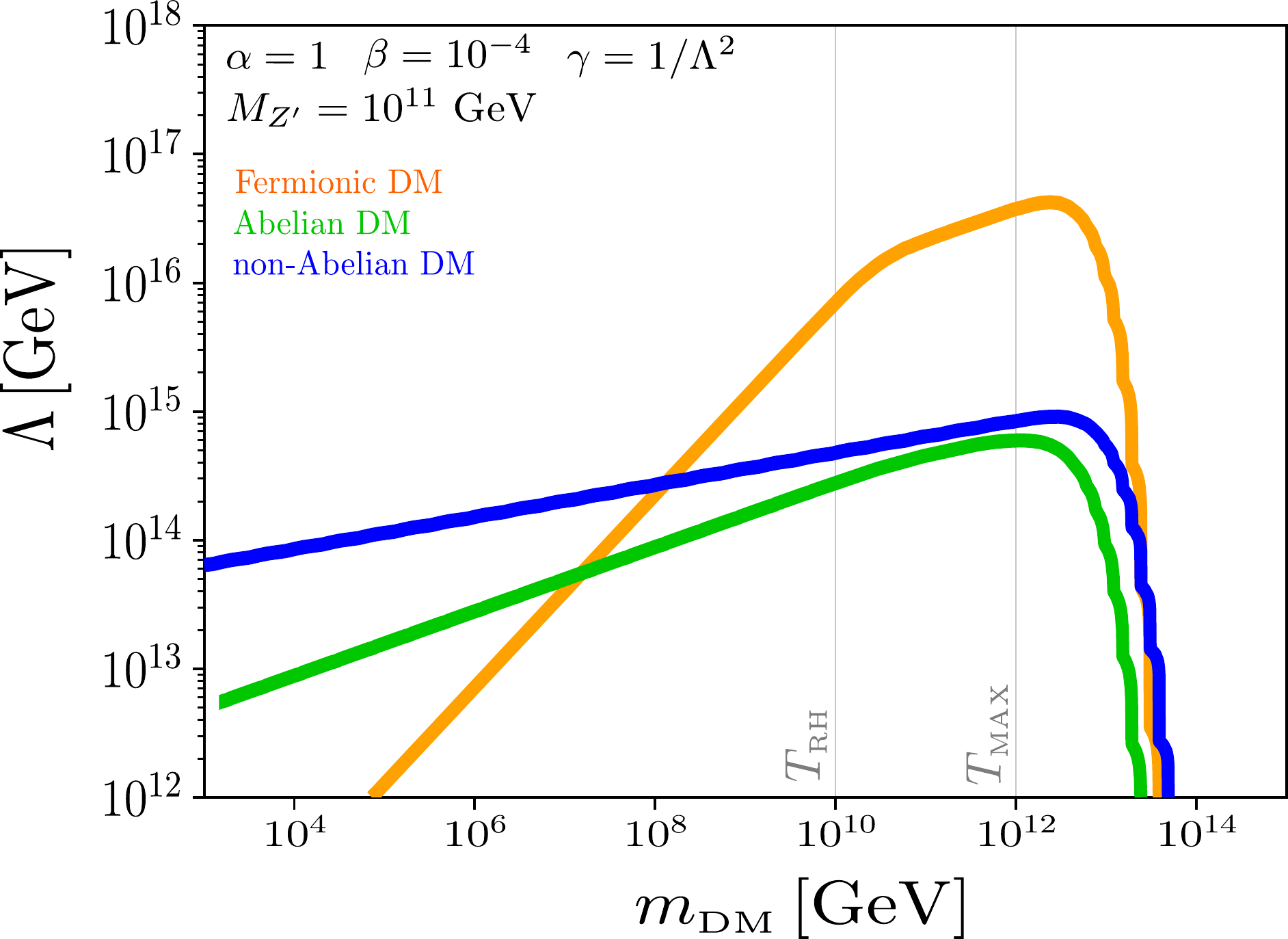}
\hspace{.3cm}
\vspace{.1cm}
\includegraphics[scale=0.3]{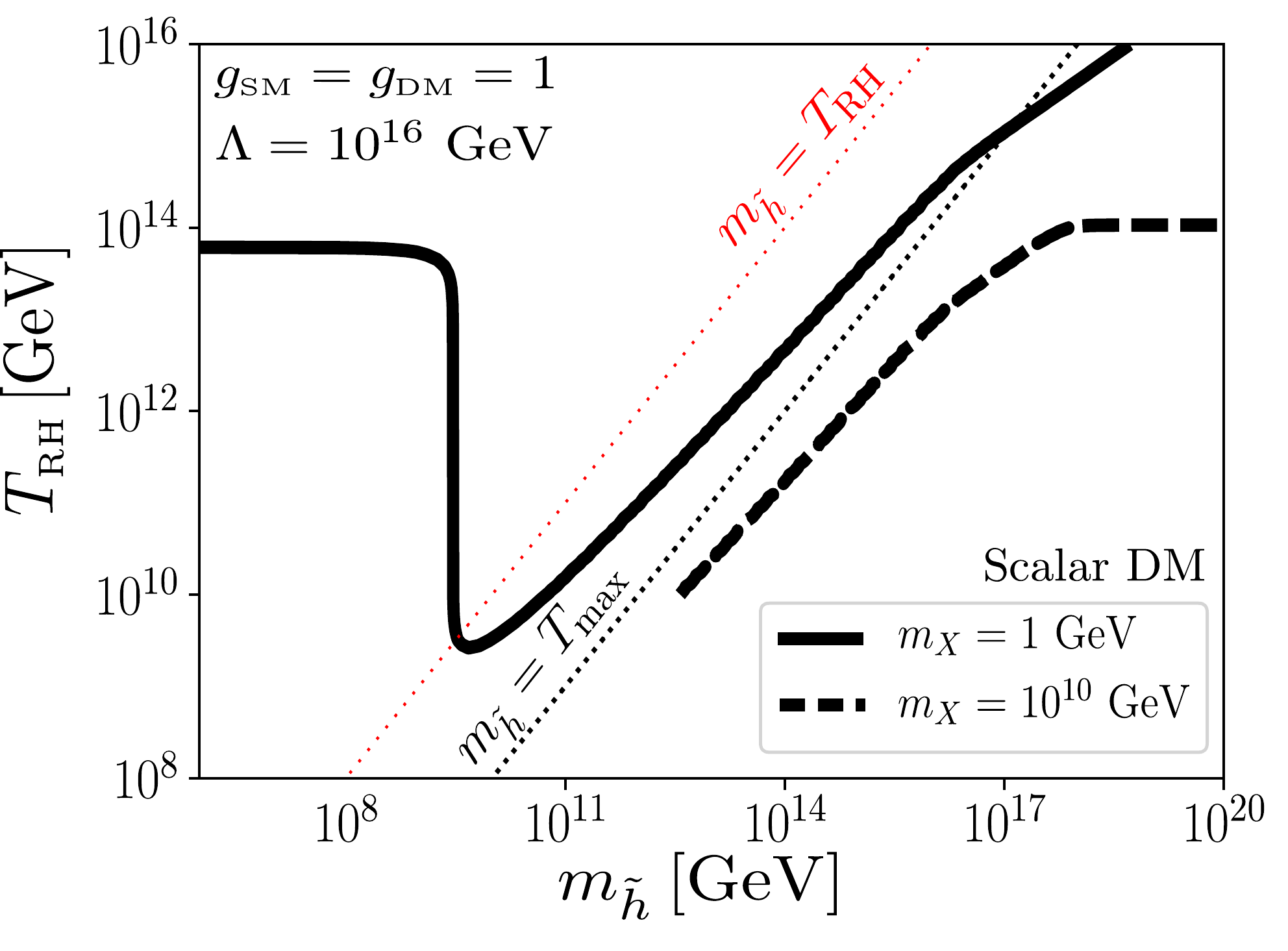}
\caption{\em \small Slices in the parameter space of the spin-1 (left) and spin-2 (right) portals in agreement with the Planck satellite constraints.}
\label{fig:slices}
\end{figure}

Let us now present the evolution of the relic density as well as its agreement with the Planck constraints \cite{aghanim_planck_nodate} when each of our FIMP candidates are to account for all the DM in the universe. In Fig. \ref{fig:evolution}, we show how the relic density of our FIMP candidates vary as a function of the inverse of $T$ in the spin-1 and spin-2 portals (left and right panels, respectively). As expected from our analysis of the T-dependence of $R_\dm$, in the spin-1 portal the freeze-in always happen before the start of radiation era, and the higher the T-dependence, the earlier their production. In the spin-2 portal, since we have a resonant exchange of $\tilde h$, the freeze-in happens either at the beginning of radiation era or whenever the $\tilde h$ pole is reached ($T \sim m_{\tilde h}$). In Fig. \ref{fig:slices}, we show the contours in our parameter spaces providing the right amount of DM today. As in all contour plots of relic density, whenever the production rates are enhanced, the overall interaction strength needs to be lowered as to provide the same relic density value. In the case of the spin-1 portal (left panel), we need larger values of $\Lambda$ to not overproduce heavier FIMPs. When the FIMPs become too heavy, $\Lambda$ needs to be lowered as to compensate for the Boltzmann suppression. In the case of the spin-2 portal (right panel), the final relic density always depend on $T_\rh$, which should be compared to $m_{\tilde h}$. When $m_{\tilde h} \ll T_\rh$, the freeze-in is independent of $m_{\tilde h}$. If $m_{\tilde h} \sim T_\rh$, $\tilde h$ is produced on-shell and then decay into DM, leading to a huge enhancement of the production rate and therefore $T_\rh$ needs to be sharply lowered to avoid overproduction. When the exchange of a too heavy $\tilde h$ becomes more than Planck-suppressed, gravitons dominate the freeze-in. When $m_{\tilde h} \sim T_\rh$, the underestimation of the relic density can be of many orders of magnitude if we use the instantaneous reheating approximation and this result depends strongly on the mediator mass and the duration of the reheating process.

\section{Conclusions}\label{conclusions}

The freeze-in production of DM in the early universe happens when their overall interaction with the SM particles is not strong enough to have led both sectors into chemical equilibrium. We have considered two scenarios in which such an out-of-equilibrium origin is inevitable -- a $Z^\prime$ and a massive spin-2 portals. The masses of these mediators are expected to be at intermediate scales of the breaking of a larger symmetry group, which might coincide with the post-inflationary reheat scale. It is therefore important to take carefully into account the finite duration of a reheating process for a correct prediction of DM relic density. In particular, this is crucial for the models considered here, where effective couplings lead to high temperature-dependencies. In both cases, SM species are able to populate the universe with the right amount of DM via freeze-in, in a wide range of DM masses (MeV to $100\, T_\max$). The prospects for detection of such DM candidates are therefore well motivated and will be explored in future works.


\begin{thebibliography}{10}

\bibitem{bhattacharyya_freezing-dark_2018}
G.~Bhattacharyya, M.~Dutra, Y.~Mambrini and M.~Pierre,
  ``Freezing-in dark matter through a heavy invisible $Z^\prime$,''
  Phys.\ Rev.\ D {\bf 98}, no. 3, 035038 (2018)
  doi:10.1103/PhysRevD.98.035038
  [arXiv:1806.00016 [hep-ph]].

\bibitem{bernal_spin-2_2018}
N.~Bernal, M.~Dutra, Y.~Mambrini, K.~Olive, M.~Peloso and M.~Pierre,
  ``Spin-2 Portal Dark Matter,''
  Phys.\ Rev.\ D {\bf 97}, no. 11, 115020 (2018)
  doi:10.1103/PhysRevD.97.115020
  [arXiv:1803.01866 [hep-ph]].

\bibitem{chung_production_1998} 
  D.~J.~H.~Chung, E.~W.~Kolb and A.~Riotto,
  ``Production of massive particles during reheating,''
  Phys.\ Rev.\ D {\bf 60}, 063504 (1999)
  doi:10.1103/PhysRevD.60.063504
  [hep-ph/9809453].

\bibitem{hall_freeze-production_2010}
  L.~J.~Hall, K.~Jedamzik, J.~March-Russell and S.~M.~West,
  ``Freeze-In Production of FIMP Dark Matter,''
  JHEP {\bf 1003}, 080 (2010)
  doi:10.1007/JHEP03(2010)080
  [arXiv:0911.1120 [hep-ph]].
  
\bibitem{bernal_dawn_2017}
 N.~Bernal, M.~Heikinheimo, T.~Tenkanen, K.~Tuominen and V.~Vaskonen,
  ``The Dawn of FIMP Dark Matter: A Review of Models and Constraints,''
  Int.\ J.\ Mod.\ Phys.\ A {\bf 32}, no. 27, 1730023 (2017)
  doi:10.1142/S0217751X1730023X
  [arXiv:1706.07442 [hep-ph]].
  
\bibitem{arcadi_waning_2018}
G.~Arcadi, M.~Dutra, P.~Ghosh, M.~Lindner, Y.~Mambrini, M.~Pierre, S.~Profumo and F.~S.~Queiroz,
  ``The waning of the WIMP? A review of models, searches, and constraints,''
  Eur.\ Phys.\ J.\ C {\bf 78}, no. 3, 203 (2018)
  doi:10.1140/epjc/s10052-018-5662-y
  [arXiv:1703.07364 [hep-ph]].
  
\bibitem{hambye_direct_2018}
T.~Hambye, M.~H.~G.~Tytgat, J.~Vandecasteele and L.~Vanderheyden,
  ``Dark matter direct detection is testing freeze-in,''
  Phys.\ Rev.\ D {\bf 98}, no. 7, 075017 (2018)
  doi:10.1103/PhysRevD.98.075017
  [arXiv:1807.05022 [hep-ph]].
  
\bibitem{giudice_largest_2001}
G.~F.~Giudice, E.~W.~Kolb and A.~Riotto,
  ``Largest temperature of the radiation era and its cosmological implications,''
  Phys.\ Rev.\ D {\bf 64}, 023508 (2001)
  doi:10.1103/PhysRevD.64.023508
  [hep-ph/0005123].
  
\bibitem{dutra_origins_2019}
 M.~Dutra,
  ``Origins for dark matter particles : from the "WIMP miracle" to the "FIMP wonder",''
  tel-02100637, 2019SACLS059.
  
  
\bibitem{anastasopoulos_anomalies_2006}
P.~Anastasopoulos, M.~Bianchi, E.~Dudas and E.~Kiritsis,
  ``Anomalies, anomalous U(1)'s and generalized Chern-Simons terms,''
  JHEP {\bf 0611}, 057 (2006)
  doi:10.1088/1126-6708/2006/11/057
  [hep-th/0605225].

\bibitem{aghanim_planck_nodate}
 N.~Aghanim {\it et al.} [Planck Collaboration],
  ``Planck 2018 results. VI. Cosmological parameters,''
  arXiv:1807.06209 [astro-ph.CO].


\end{thebibliography}
\end{document}